\documentclass[aps,prl,twocolumn]{revtex4-1}

\pdfoutput=1
\usepackage[final]{graphicx}
\usepackage[utf8]{inputenc}
\usepackage{amssymb,amsmath,amsbsy,amsthm,mathrsfs}
\usepackage[USenglish]{babel}
\usepackage{hyphenat}
\usepackage[hyperref]{xcolor}
\usepackage{bm}
\usepackage{subfigure}
\usepackage{enumerate}
\usepackage{units}
\usepackage{textcomp}
\usepackage{ulem}
\usepackage{xcolor}
\usepackage{graphicx}
\usepackage{epstopdf}
\usepackage{xr}
\usepackage[breaklinks=true,
            pdfborder={0 0 1},
            colorlinks=true,
            linkcolor=black,
            citecolor=blue,
            urlcolor=blue]{hyperref}

\newcommand{\ud}{\mathrm{d}}

\newcommand{\abs}[1]{\left\lvert #1 \right\rvert}

\begin{document}


\title{Dynamics of growth and form in prebiotic vesicles}


\author{Teresa Ruiz-Herrero}
\affiliation{John A. Paulson School of Engineering and Applied Sciences, Harvard University, Cambridge, Massachusetts 02138,USA}
\author{Thomas G. Fai}
\affiliation{Department of Mathematics, Brandeis University, Waltham, Massachusetts 02453,USA} 
\author{L. Mahadevan}
 \affiliation{John A. Paulson School of Engineering and Applied Sciences,  Department of Physics, Department of Organismic and Evolutionary Biology, Harvard University, Cambridge, Massachusetts 02138,USA }



\begin{abstract}

The growth, form, and division of prebiotic vesicles, membraneous bags of fluid of varying components and shapes is hypothesized to  have served as the substrate for the origin of life. The dynamics of these out-of-equilibrium structures is controlled by physicochemical processes that include the intercalation of amphiphiles into the membrane, fluid flow across the membrane, and elastic deformations of the membrane. To understand prebiotic vesicular forms and their dynamics, we construct a minimal model that couples membrane growth, deformation, and fluid permeation, ultimately couched in terms of two dimensionless parameters that characterize the relative rate of membrane growth and the membrane permeability. Numerical simulations show that our model captures the morphological diversity seen in extant precursor mimics of cellular life, and might provide simple guidelines for the synthesis of these complex shapes from simple ingredients. 
  
\end{abstract}


\maketitle

\begin{figure}
 \centering

\includegraphics[bb=0 0 936 720, width=0.45\textwidth]{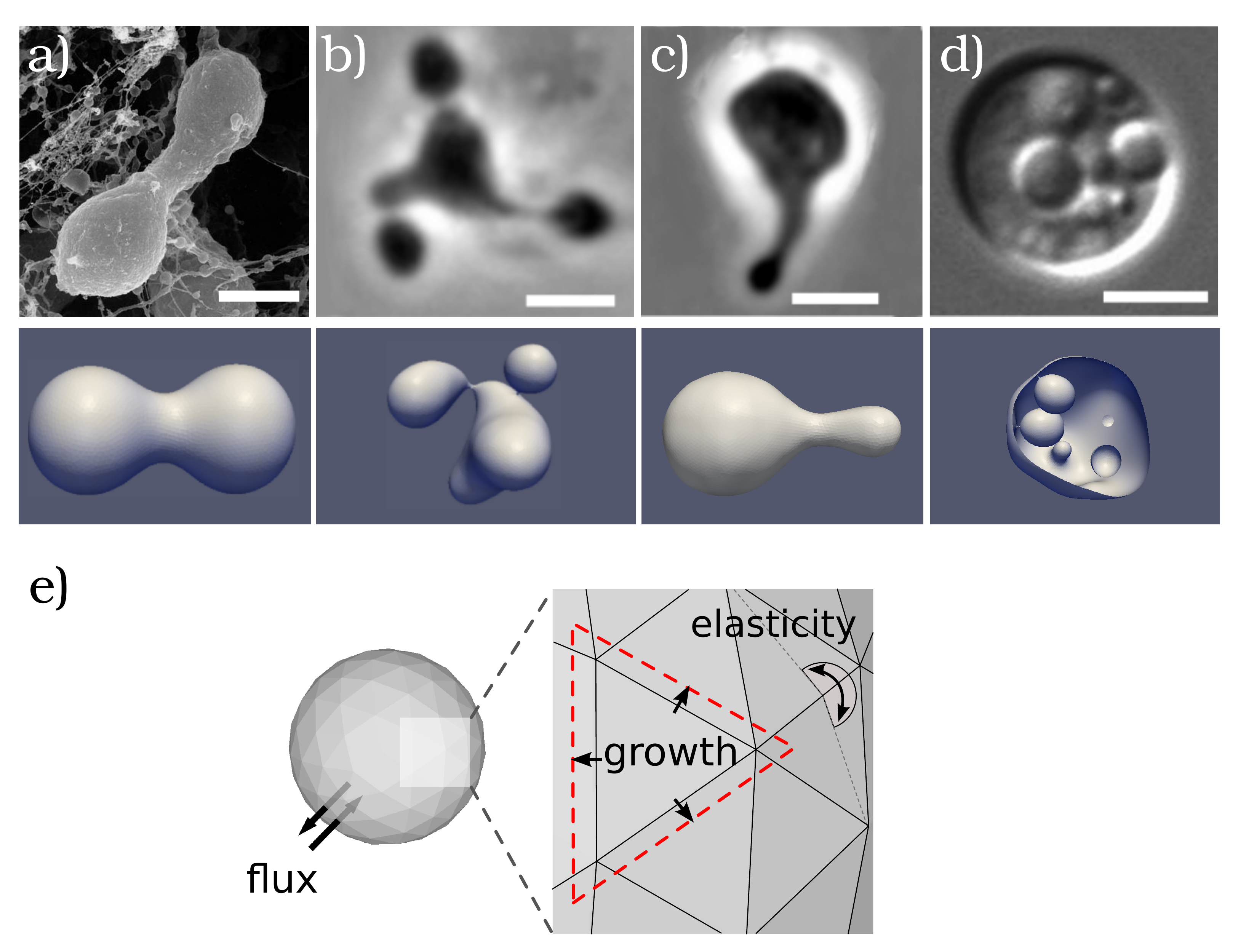}

  \caption {(top) Different morphologies observed during growth in synthetic giant vesicles and L-form bacteria: a) symmetric division, adapted from \cite{Onoda2000} b) budding, adapted from Refs. \cite{Mercier2013b, Bendezu2008} c) tubulation, adapted from  \cite{Mercier2013b} d) vesiculation, adapted from  \cite{Mercier2013b, Bendezu2008}  (Scale bars represent $3\mu m$). Our minimal model leads to shapes that are similar to those observed experimentally, using the following dimensionless parameters: a) $\Pi_1 = 0.01$,  $\Pi_2 = 2.5$ , b) $\Pi_1 = 0.15$,  $\Pi_2 = 5$, c) $\Pi_1 = 0.02$,  $\Pi_2 = 5$ and d) $\Pi_1 = 0.15$,  $\Pi_2 = -2.5$ (see text for details).  e) A schematic of our vesicle model that uses a 3D triangulated lattice with bending rigidity $B$ immersed in a fluid with viscosity $\mu$, with area that grows with homogeneous expansion of the triangle size at a rate $\gamma$ and volume whose evolution is controlled by the wall permeability $K$.  }
\label{fig:intro}
 \end{figure}

It is likely that the first cells originated when a self-replicating biomolecule was separated from its environment by a permeable membrane barrier and both the biomolecule and the membrane were able to grow and replicate. Physical compartmentalization allowed for a separation of chemical environments, making way eventually for the specialization and competition between cells that is the basis for Darwinian evolution \cite{Budin2010,Sole2009}. How these prebiotic cells could grow and divide without the complex machinery in extant cells remains a major open question in biology. Given the strong chemical and physical constraints on biomolecular replication, and membrane compartmentalization, growth, and dynamics, it is natural to expect that physicochemical processes are intimately tied to the evolvability of such states. Recent research on the ability of a biomolecule to replicate and transmit information has led to a consensus on a range of possible chemical replicators \cite{Higgs2014}. Independently, the physical properties of the external membrane barrier under growth and division have also been the subject of experimental studies \cite{Zhu2009,Budin2011}. However, the phase space of physical solutions for the growth and form of the prebiotic vesicles is difficult to grapple with owing to the range of spatio-temporal processes that need to be accounted for---from membrane growth and deformation to fluid permeation and ultimately division. Insight into the dynamics of membrane growth and replication may be gleaned by considering artificial lipid vesicles as well as naturally occurring L-form bacteria. Synthetic lipid vesicles composed of single-chain amphiphiles are considered to be representative of prebiotic conditions \cite{Chen2004b}, as are L-forms, naturally occurring bacteria with genetic mutations that inhibit cell wall formation \cite{Errington2013}. Both these systems have been experimentally shown to exhibit complex shapes and modes of growth; they can grow while maintaining their original spherical shape, by elongating into cigar shapes that eventually divide into two vesicles of the same size \cite{Mercier2012a, Leaver2009, Bendezu2008, Onoda2000} (Fig.~\ref{fig:intro}a), or by developing protrusions in the form of external buds \cite{Leaver2009,Peterlin2009, Mercier2013b} (Fig.~\ref{fig:intro}b), internal buds \cite{Bendezu2008, Peterlin2009} (Fig.~\ref{fig:intro}d), or long tubes \cite{Zhu2009, Szostak2011, Mercier2013b,Leaver2009, Peterlin2009} (Fig.~\ref{fig:intro}c). It has been suggested previously that growth and division may be controlled solely by the physical processes at play \cite{Briers2012,Budin2011,zwicker2017}. In particular, it is well-established that deformations during growth involve dynamical imbalances in the surface area to volume ratio, either due to excess membrane growth or low permeability \cite{Mercier2013b, Kaes1991,Harris2014}. Our work builds on existing theoretical studies of the equilibrium shapes of vesicles \cite{Svetina1989,Seifert1997a,markvoort2006,Kaoui2009,Sakashita2012} and out-of-equilibrium membrane growth \cite{Bozic2007, Macia2007a,Macia2007b,markvoort2007,markvoort2009,markvoort2010} by exploring the range of possible behaviors within a non-equilibrium physical model that couples membrane growth and fluid permeation.   
  
Our minimal model of prebiotic vesicles assumes a closed elastic surface of initial radius $R_0$, spontaneous curvature $c_0$, bending stiffness $B$, and fluid permeability $K$, with the membrane thickness being much smaller than the vesicle radius, which changes over time. We also assume the membrane to be nearly inextensible, which translates into a high energy cost for stretching, making bending deformations energetically preferable. The vesicle is assumed to be immersed in an incompressible fluid having viscosity $\mu$ and at temperature $T$. We further assume that  the amphiphilic molecules that constitute the membrane are at a constant concentration in the surrounding medium and that they are incorporated into the membrane at an average net rate of $\gamma$.  
 At a continuum level, this implies that the vesicle area $A$ grows according to the simple law
\begin{equation}
\dot{A}=\gamma A
\end{equation}
Since lipids are incorporated into the external layer, we assume that rapid transbilayer lipid exchange distributes amphiphiles across the membrane and relax the bending energy \cite{bruckner2009flip}, and further that amphiphile species determines the preferred spontaneous curvature \cite{sakuma2015vesicles}.
We account for the vesicle permeability, with changes in the vesicle volume produced by transmembrane fluid flux according to 
\begin{equation}
\dot{V}=AK\Delta P
\end{equation}
 where the pressure drop $\Delta P=P_\text{out}-P_\text{in}$, and $K$ is the membrane permeability.  In this minimal model, we assume that the pressure drop is dominated by the osmotic component, which is kept constant by implicit internal mechanisms. {We note that these two equations are incompatible with spherical vesicle growth, since they specify two laws for radial expansion - one linear and another exponential. Naturally, the slower of these is rate-limiting, and this leads to the complexity of shapes seen, as we will see shortly.}

Since the size of the system ($\sim 10 \mu m$)) is larger than the scale over which thermodynamic fluctuations are relevant (and $B/k_BT \sim 10$), we neglect the role of thermal fluctuations. In terms of the five variables, the bending stiffness ($B$), growth rate ($\gamma$), dynamic viscosity ($\mu$), effective permeability ($K\Delta P$), and spontaneous curvature ($c_0$), we construct three relevant length scales:  {the critical radius $R_\text{i}=K\Delta P/\gamma$, i.e.~the radius below which vesicle growth is dominated by volume increase and above which is dominated by area growth}, the mechanical relaxation lengthscale $R_\text{x}=(B/\gamma\mu)^{1/3}$, the size below which bending deformations are mechanically equilibrated, but above which they are still dynamically varying, and the lengthscale related to the spontaneous curvature $c_0^{-1}$. Using the following values for the viscosity  $\mu=0.8\cdot 10^{-3} \,\text{kg}/\text{m}\cdot\text{s}$, bending stiffness $B=10 \,k_\text{B}T=4\cdot 10^{-20}\,\text{J}$ \cite{Rawicz2000}, scaled permeability $K\Delta P=10^{-7}$--$10^{-5}\,\text{m/s}$ \cite{Sacerdote2005,Olbrich2000, Huster1997,Jansen1995}, growth rate $\gamma=0.5\, \text{s}^{-1}$ \cite{Chen2004b}, and spontaneous curvature $\abs{c_0} =10^6$--$10^8\,\text{m}^{-1}$ \cite{Kamal2009}, we find that $R_\text{i} \sim [10^{-7} - 10^{-5}] \,\text{m}$, $R_\text{x} \sim [5 \times 10^{-8}, 20 \times 10^{-7}] \,\text{m}$.  This allows us to define two dimensionless parameters: $\Pi_1=R_i/R_x \in [0.01,1]$, which accounts for the ability of the vesicle to mechanically equilibrate under imbalances arising from growth beyond $R_i$, and $\Pi_2=R_\text{i}c_0 \in [0.1,100]$, which determines the relative magnitude of spontaneous vesicle curvature (noting that it can be negative or positive). A small value of $\Pi_1$ corresponds to a small critical radius and large relaxation lengthscales:  {this is the limit of slow growth in which vesicles evolve in a sequence of quasi-equilibrated shapes. On the other hand, large values of $\Pi_1$ correspond to large critical radii and small relaxation lengthscales which allow only for local equilibration; this is the limit of non-equilibrium growth. The subset of values we consider corresponds to the regime of membrane-driven growth, which we reason is likely when simple cellular precursors are unlikely to have been able to sustain high osmotic pressures. }

 
 We use overdamped dynamics to model the vesicle as a porous elastic membrane immersed in an incompressible fluid. The elastic energy of the lipid bilayer is assumed to be equal to the sum of the local stretching energy, the Canham-Helfrich Hamiltonian \cite{can,Zhong-can1989}, and a penalty term that tethers the volume of the vesicle to the target volume $V_\text{T}$, which is typically growing. This yields the expression for the energy
 \begin{equation}
 E = \frac{k_\text{a}}{2} \int_{\mathbf{S}'} \left(J-1\right)^2 \ud a' +\frac{B}{2} \int_\mathbf{S} (H-c_0)^2 \ud a + \frac{k_\text{V}}{2}(V-V_\text{T})^2,
 \end{equation}
where $k_\text{a}$ is the stretching coefficient, $B$ is the bending modulus, $H$ is the sum of the principal curvatures, $c_0$ is the spontaneous curvature, and $k_\text{V}$ is a volume-preserving penalty parameter. In the above integrals, $\ud a'$ is the area element in the reference surface $\mathbf{S}'$,  $\ud a$ is the area element in the deformed configuration $\mathbf{S}$, and the term $J$ that appears in the stretching energy is the Jacobian of the transformation from reference coordinates to deformed coordinates. The reference surface, i.e.~the equilibrium state, is assumed to be a sphere, with the stretching term penalizing local changes in area relative to the reference configuration following (1), while the target volume follows (2). In our simulations, the membrane is represented as a triangulated lattice that undergoes growth and deformation (Fig.~\ref{fig:intro}e), with vertices following Brownian dynamics in the presence of forces driven by the Hamiltonian above. To avoid numerical instabilities, the surface is remeshed periodically and the effective temperature is kept very small to ensure robustness with respect to mesh size and shape changes, and small fluctuations (see SI for further details).

\begin{figure}
 \centering
\hspace{-0.25in}\includegraphics[bb=0 0 1200 960,width=0.5\textwidth]{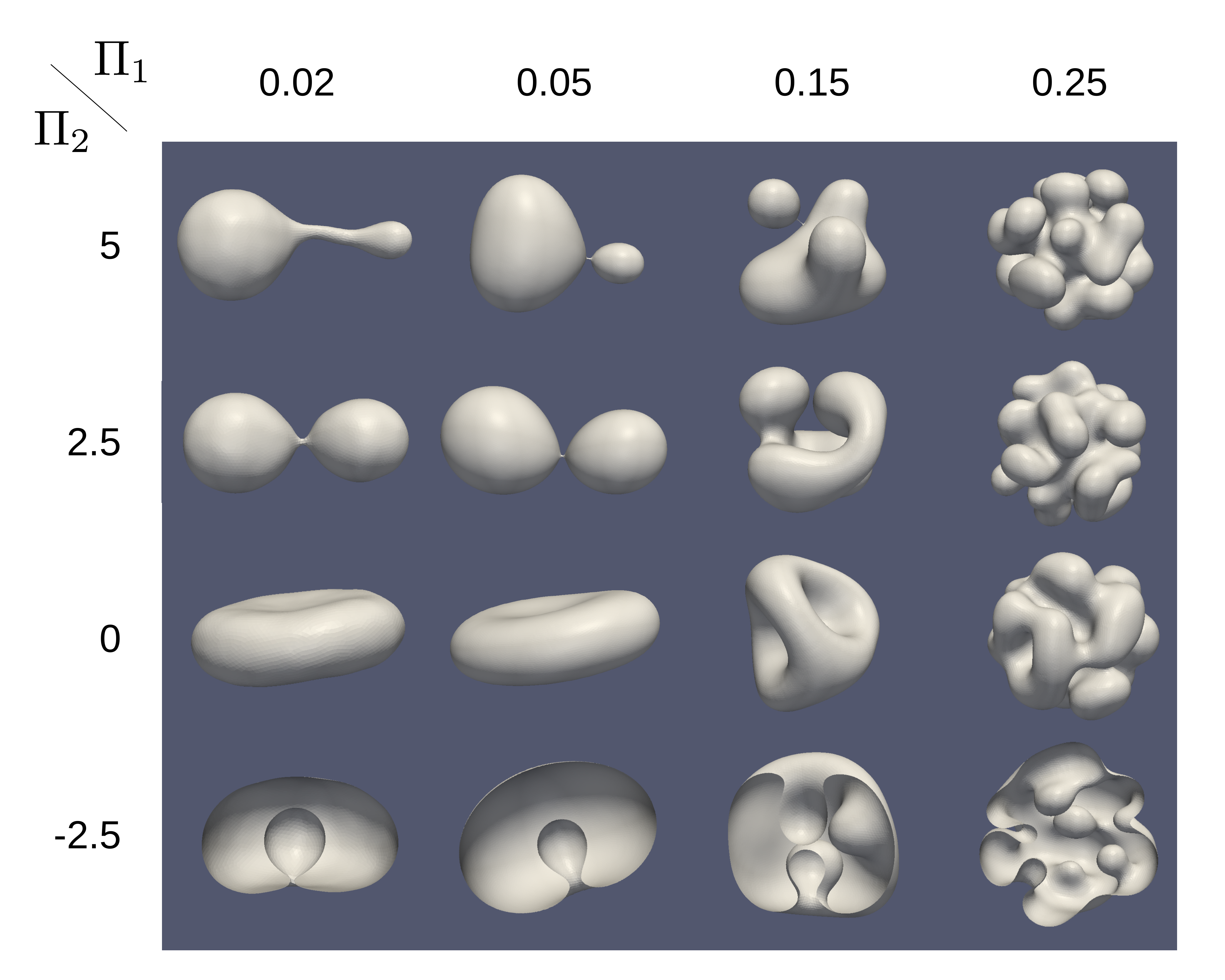}

  \caption {Morphospace of vesicle shapes as a function of the dimensionless mechanical relaxation $\Pi_1$ and the dimensionless spontaneous curvature $\Pi_2$. For $\Pi_2=-2.5$ the shapes also visualize the interior of the vesicles where vesiculation occurs.  {Configurations correspond to vesicle shapes immediately prior to division or fission, with snapshots of the vesicle just before a topological transition associated with fission.}}
\label{fig:3D}
 \end{figure}

\begin{figure}
 \centering
 
 \hspace{-0.2in}\includegraphics[bb=0 0 2080 2096,width=0.5\textwidth]{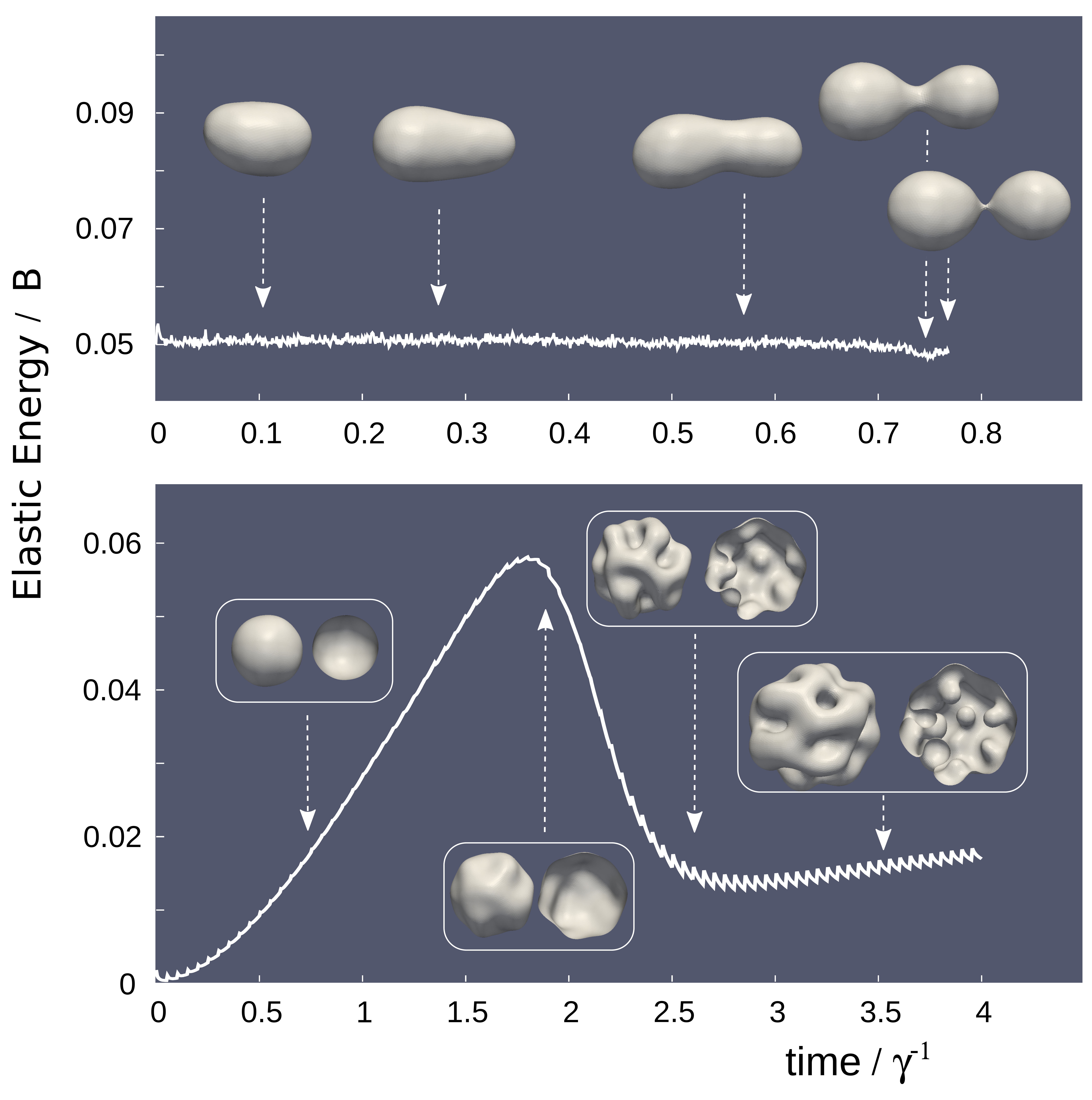}

  \caption {Scaled elastic energy and vesicle shapes  as a function of scaled time for two different modes of growth. (Top)  When $\Pi_1 = 0.05$ and $\Pi_2 =2.5$, the formation of a skinny neck between symmetric lobes provides a likely mechanism of homeostatic division in 3D. The slow growth allow the vesicle to deform through quasi-equilibrated shapes with roughly constant elastic energy, and with a drop in the energy corresponding to neck formation. (Bottom) When $\Pi_1=0.25$ and  $\Pi_2=-2.5$, fast surface growth and negative curvature lead to multiple sites of inward vesiculation. The build-up in elastic energy is a signature of fast non-equilibrated growth. The shapes along the curve also show the interior of the vesicles.}
\label{fig:division}
 \end{figure}

We simulated vesicular growth using this model after initializing the vesicles as spheres with initial radius $R_0=2R_i$ over the range $\Pi_1=0.01$--$0.5$ and $\Pi_2=-2.5$--$5$ by varying the growth rate, permeability, bending stiffness, and viscosity. First we study the shape evolution during growth as a function of $\Pi_1$ for vesicles with zero spontaneous curvature ($\Pi_2=0$ corresponding to the intermediate row of Fig.~\ref{fig:3D}). In all our simulations reported in the paper, we have chosen the stretching coefficient $k_\text{a}$ to be sufficiently large so that bending, rather than in-plane stretching, is the preferred mode of deformation. We find a transition that occurs continuously around $\Pi_1=0.15$ with shapes showing increasingly high-order symmetries. Values of $\Pi_1$ below this transition correspond to quasi-equilibrum shapes that continuously relax while the reduced volume decreases during growth (Fig.~\ref{fig:division} (a)). Values of $\Pi_1$ above the transition correspond to nonequilibrated configurations in which surface growth is faster than the timescale for mechanical relaxation, so that the vesicle incorporates new material by corrugating its surface at the cost of increased elastic energy.
 
For the case of zero spontaneous curvature ($\Pi_2=0$), there is an energy barrier for neck formation that prevents budding or sprouting. Consequently, in the quasi-equilibrated case the growing surface area can only be accommodated by the formation of vesicle-scale, pancake-like geometries. The most general way to form necks and thus take the simplest route to cell division, is by introducing a non-zero spontaneous curvature. Indeed, for fixed non-zero spontaneous curvature, with $\Pi_2 \ne 0$, we see the emergence of two different behaviors depending on the sign of the spontaneous curvature. Positive spontaneous curvatures give rise to tube formation and budding. Consistent with the observations for $\Pi_2=0$, we observe quasi-equilibrium shapes at small values of $\Pi_1$ in which a tube sprouts from the main body of the vesicle. As $\Pi_1$ is increased, tube formation is replaced by single budding events. At large values of $\Pi_1$, several budding sites emerge on the vesicle surface. Finally, negative spontaneous curvature corresponds to shapes with inner tubulation (small $\Pi_1$) and inner vesiculation (large $\Pi_1$, Fig.~\ref{fig:division}(b)). In the SI, we investigate the case of low $k_\text{a}$ in which surface stretching becomes energetically preferable and find that rather than tube sprouting, a neck appears in the narrowest section of a pear-shaped vesicle (Fig. S1).
 
Large values of $\Pi_1$ correspond to the cases of high permeability and rapid growth, in which both vesicle volume and surface area grow faster than the timescale for mechanical relaxation, resulting in a build-up of elastic energy (Fig. ~\ref{fig:division}(b)). The vesicle grows spherically until volume growth cannot keep up with surface growth, at which point patches of constant mean curvature with $\abs{c}\sim c_0$ appear throughout the surface to relax the bending energy. Further surface growth results in the accumulation of extra material in those patches, which subsequently become nucleation sites for budding or vesiculation.
 
Although we stop our simulations prior to vesicle fusion or division given the geometric and biophysical complexity of the topological transition associated with division in 3 dimensions, we can explore this process in the case of 2 dimensions (relevant for vesicles that are confined between solid surfaces) and also study the formation of thin necks Fig.~\ref{fig:division}(a)), since this might lead to division spontaneously due to thermal fluctuations. Our qualitative exploration shown in Fig. 2 reveals various behaviors: we find vesicles approaching symmetric division with very small dispersion in size, and vesicles that develop small internal or external buds that might also be precursors to division. In this context, it is important to note that the initial radius influences the shape of the vesicles during growth. Assuming a spherical configuration and setting the radius change from the area and volume growth equations equal to each other, one may compute $R_c=2K \Delta P /\gamma=2R_i$ to be the radius at which volume growth cannot keep up with surface growth and the vesicle begins to deviate from a spherical shape. Whereas the above results were obtained using an initial vesicle radius of $R_0=R_c$, if $R_0<R_i$, there is a preliminary stage in which the vesicle grows spherically until reaching the critical radius $R_c$ before the deformations discussed above occur.  {When $R_0>R_c$ however, area growth is initially much faster than volume growth and the surface undergoes corrugations at lower values of $\Pi_1$; for large values of $R_0$, daughter vesicles will effectively bud off, reducing the radius of the mother vesicle until it reaches $R_c$.}

Simulations in 2D systems (see SI), show very similar features that map onto the morphospace of Fig.~\ref{fig:3D}, in the sense that vesicles will exhibit the morphologies of Fig.~\ref{fig:3D}  immediately prior to division. Furthermore, in the 2D systems, we can capture the topological transitions associated with division easily and thus simulate multiple generations (see SI). Vesicles that grow into cigar shapes display accurate size control when the permeability and growth rate are such that both the perimeter and area double simultaneously (SI Fig.~S5), leading to a periodic steady state.

To assess the validity of our assumption of local hydrodynamics, we used the immersed boundary method \cite{acta} to model the non-local hydrodynamics and solved the fully coupled elastohydrodynamic problem (see SI). While our results are qualitatively consistent with the simpler local hydrodynamic approximation used so far, accounting for non-local hydrodynamics increases the characteristic length scales of membrane tubules and invaginations and lowers the energy barrier for the formation of creases and folds (see SI). 

Overall, our study of non-equilibrium vesicle growth and division allows us to investigate the role of permeability, stiffness, viscosity, and growth rate via two dimensionless parameters that define a two-dimensional morphospace. Our simulations reveal that many of the essential aspects of growth and dynamics can be understood in terms of an imbalance between surface to volume growth and the relative rate of mechanical relaxation. Our morphospace allows us to recapitulate the various observed shapes of simple dynamically growing lipid vesicles and their approximate biological analogs, L-forms \cite{Mercier2012a, Leaver2009, Bendezu2008, Onoda2000, Peterlin2009}, and allows us to evaluate whether the varied morphodynamics of prebiotic vesicles and their modern counterparts could arise from non-equilibrium physicochemical processes. Our minimal model provides a foundation to study the physicochemical constraints on protocellular growth and replication while setting the stage to include the additional complexity associated with the dynamics of transbilayer lipid exchange and natural curvature, internal sources of lipids, concentration differences across the membrane, and the role of multiple bilayers.

\begin{acknowledgments}
We thank Chris Rycroft, Charles Peskin, Chen-Hung Wu, and Etienne Vouga for useful discussions, and the Ramón Areces Foundation (TRH), the National Science Foundation grant DMS-1502851 (TGF), and the Simons Collaboration on the Origins of Life grant 390237 (TRH and LM) for partial support.
\end{acknowledgments}


Refs.~\cite{Kim2006, Kim2012, Kim2014,lilai,vc_ib,vc_ib2,brakke1992surface,Botsch2010,Kooijman2005,Weeks1971} appear in the Supplemental Information.

\end{document}


\title{Supplementary Information for \\``Dynamics of growth and form in prebiotic vesicles"}
 \author{Teresa Ruiz-Herrero}
\affiliation{School of Engineering and Applied Sciences, Harvard University, Cambridge, Massachusetts 02138,USA}
\author{Thomas G. Fai}
\affiliation{Department of Mathematics, Brandeis University, Waltham, Massachusetts 02453,USA} 
\author{L. Mahadevan}
 \affiliation{School of Engineering and Applied Sciences, Harvard University, Cambridge, Massachusetts 02138,USA}
\affiliation{Department of Physics, Department of Organismic and Evolutionary Biology, Harvard University, Cambridge, Massachusetts 02138,USA }
\maketitle

\section{3D Model}

 \subsection{Local hydrodynamic formulation}
 
 The algorithms for the local elastohydrodynamic formulation are given in the Main Text. Here, we show the results of a simulation designed to study the effect of the stretching coefficient. 
 \begin{figure}[H]
  \centering
 \includegraphics[bb=0 0 1440 1040,width=0.6\textwidth]{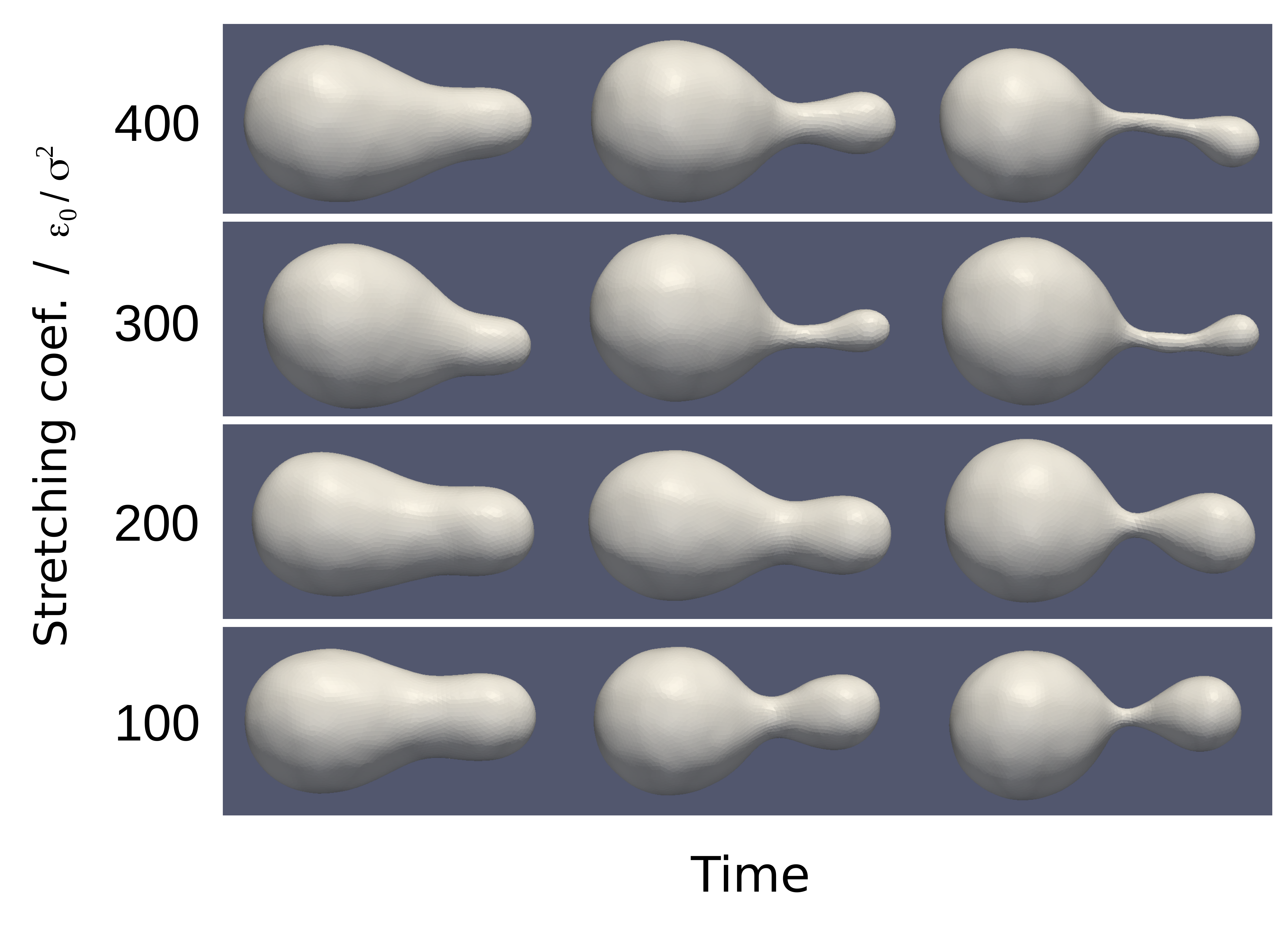}
\caption {{Effect of the stretching coefficient $k_\text{a}$ on vesicle growth using local hydrodynamics. Each row shows the shape progression for a particular value of $k_\text{a}$. Tube sprouting only occurs in the non-stretching regime of large $k_\text{a}$. We take $\Pi_1=0.02$ and $\Pi_2=5$ to be fixed. Note that trajectories evolve at different speeds so that vesicle configurations along columns do not correspond to snapshots taken at the same time.}}
\label{fig:SI-Figtubes}
 \end{figure}
 
 \subsection{Non-local hydrodynamic formulation}
 
As mentioned in the Main Text, in addition to the overdamped simulations we have used the immersed boundary method \cite{acta} to simulate the hydrodynamics of growing poroelastic vesicles immersed in fluid. Growth is assumed to be homogeneous and simulations are stopped prior to any division or fusion events. The vesicle is parameterized by Lagrangian coordinates $\mb{q}=(q,r)$, and the Cartesian position of the vesicle at time $t$ is given by the function $\mb{X}(\mb{q},t)$. The fluid surrounding the vesicle is modeled explicitly by the incompressible Navier-Stokes equations with fluid velocity $\mb{u}$ and pressure $p$. The fluid and elastic material are coupled as follows: the configuration $\mb{X}(\mb{q},t)$ gives rise to a Lagrangian force density $\mb{F}(\mb{q},t)$, which is transmitted to the fluid as a delta-function layer of force supported on the vesicle surface. Further, following previous authors \cite{Kim2006,Kim2012,Kim2014} and incorporating an additional osmotic pressure term, we assume the elastic material moves at a velocity given by
\begin{align}
 \frac{\partial \mb{X}}{\partial t}(\mb{q},t) = \mb{U}(\mb{q},t)+K\left(\nabla \phi+ \frac{\left(\mb{F}(\mb{q},t)\cdot\mb{N}(\mb{q},t)\right)\mb{N}(\mb{q},t)}{\norm{\frac{\partial \mb{X}}{\partial r} \times \frac{\partial \mb{X}}{\partial s}}}\right), \label{ib:perm}
\end{align}
where $\mb{U}(\mb{q},t)=\mb{u}(\mb{X}(\mb{q},t),t)$ is the fluid velocity, $\mb{N}$ is the unit normal to the vesicle, $\nabla \phi$ is the applied osmotic pressure, and $K$ is the permeability. This is equivalent to having a local flux across the membrane proportional to the jump in pressure \cite{lilai}. As mentioned in the Main Text, if $K = 0$, the vesicle moves at the local fluid velocity (i.e.\ the no-slip condition is satisfied) and volume is conserved. For nonzero $K > 0$, the membrane is porous, allowing relative slip between the fluid and vesicle membrane, and the enclosed volume increases over time.

Together with \eqref{ib:perm}, the continuous immersed boundary formulation consists of the following system of equations for $\mb{u}$, $p$, and $\mb{X}$:
\begin{align}
\rho \left(\frac{\partial \mathbf{u}}{\partial t}+\mathbf{u} \cdot \nabla\mathbf{u}\right)+\nabla p
&=\mu \Delta \mathbf{u}+ \mathbf{f} \label{ib:mom}\\
\nabla \cdot \mathbf{u} &= 0 \\
\mathbf{f}(\mathbf{x},t)&= \int \mathbf{F}(\mathbf{q},t) \delta \left( \mathbf{x}-\mathbf{X}(\mathbf{q},t) \right)\ud \mathbf{q} \label{ib:spread} \\
\mathbf{U} (\mathbf{q},t)&= \int \mathbf{u}(\mathbf{x},t) \delta \left( \mathbf{x}-\mathbf{X}(\mathbf{q},t) \right) \ud \mathbf{x} \label{ib:interp} \\
\mathbf{F} &=-\frac{\delta E}{\delta \mathbf{X}}, \label{ib:en}
\end{align}
where $\delta E/\delta \mathbf{X}$ represents the variational derivative of the elastic energy. An elastic energy functional $E[\mathbf{X}(\cdot,t)]$ must be specified to determine the Lagrangian force density via \eqref{ib:en}.
 {
The elastic energy $E$ in this model is similar to Main Text Eq.~(1), with bending rigidity $B$, spontaneous curvature $c_0$, and local stretching resistance $k_\text{a}$:
\begin{equation}
 E = \frac{k_\text{a}}{2} \int_{\mathbf{S}'} \left(J-1\right)^2 \ud a' +\frac{B}{2} \int_\mathbf{S} (H-c_0)^2 \ud a. \label{ib:energy}
\end{equation}
As before, $H$ is the sum of the principal curvatures, $\ud a'$ is the area element in the reference surface $\mathbf{S}'$, $\ud a$ is the area element in the deformed configuration $\mathbf{S}$, and $J$ is the Jacobian of the transformation from reference coordinates to deformed coordinates. Growth in surface area is implemented by increasing the reference area at a prescribed rate $\gamma$ according to $\dot{A} = \gamma A$. As the reference area increases the membrane is placed under compression and equilibrates on expansion. Unlike the local hydrodynamics model, the incompressibility constraint (S.3) used here ensures that volumes are preserved. In practice, in the zero-permeability case we observe that vesicle volumes change by less than $1\%$ over the course of our simulations. Therefore the volume-preserving penalty parameter appearing in Main Text Eq.~(1) is not needed in the full hydrodynamics simulations.}

 {
To solve the system of equations \eqref{ib:perm}-\eqref{ib:en} using the immersed boundary method, the fluid domain is discretized using a uniform grid and the vesicle is discretized using the same triangulated surface as above. Periodic boundary conditions are imposed on the fluid domain, and the fluid equations are solved subject to this forcing using the fast Fourier transform. The fluid velocity is interpolated to the elastic material by discretizing \eqref{ib:interp}, and the material is advanced to the next timestep using velocity given by \eqref{ib:perm}. We continue in this manner, advancing from timestep to timestep until the prescribed endtime is reached. See \cite{vc_ib,vc_ib2} for further implementation details, including a discussion about the discretization on triangulated surfaces of Lagrangian quantities such as curvature \footnote{The curvature discretization used here has the same formulation as the method {\texttt{star\_perp\_sq\_mean\_curvature}} of Surface Evolver \cite{brakke1992surface}.}.
} Our simulation protocol is described below.

\begin{enumerate}
\item Start simulation with triangulated spherical mesh of initial radius $R_0=R_\text{c}$.
\item For each timestep:
  \begin{enumerate}
  \item Perform force calculation 
  \item Update position and velocity
  \item Update the reference area  of each element of the mesh $a'$  following $a'(t) = a'(t-1) + a(t-1)\gamma dt$, where $a$ is the actual area of each element in the mesh.   
  \item Update the target volume according to $V_\text{T}(t) = V_\text{T}(t-1) + A(t)K\Delta P dt$, where $A$ is the total area of the vesicle.
  \end{enumerate}

\item For every $n_\text{remesh}$ steps
  \begin{enumerate}
  \item Remesh (see below for more information)
  \end{enumerate}

\item Terminate when material from opposite sides of the vesicle comes close to touching or a thin neck is formed.

\end{enumerate}
 
\begin{figure}[h!]	
  \includegraphics[bb=0 0 1680 560,width=0.95\textwidth]{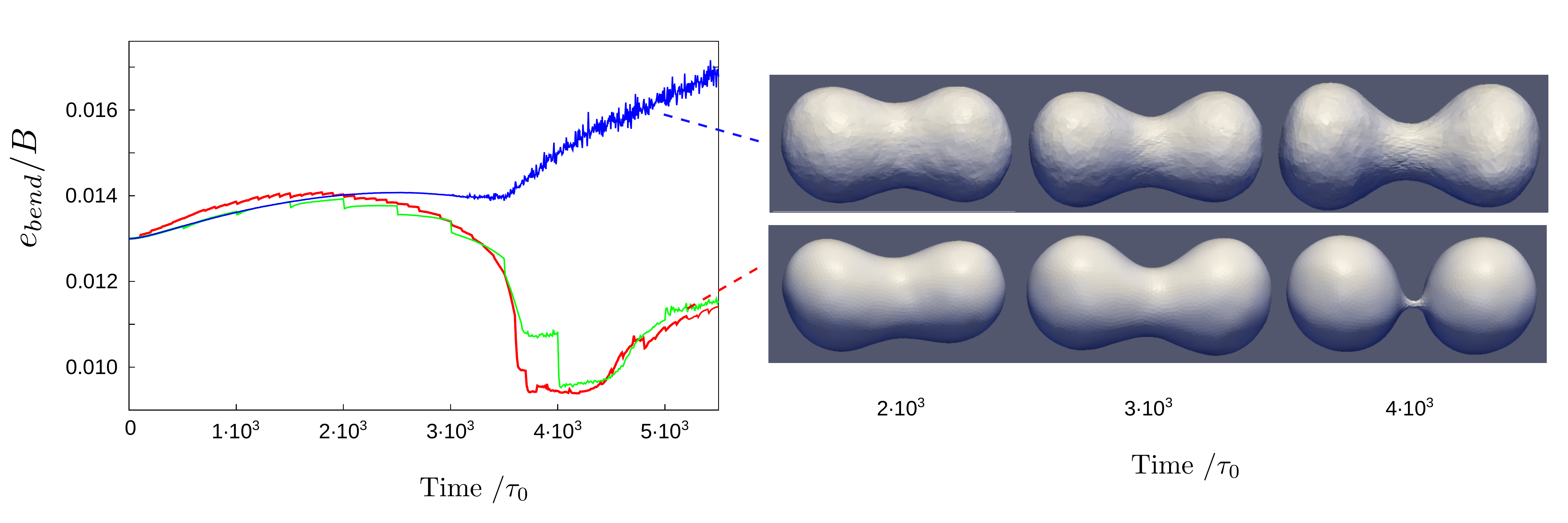}

  \caption{Effect of remeshing frequency during vesicle growth. On the left is shown the bending energy per vertex $e_\text{bend}$, scaled by bending stiffness, for three remeshing frequencies: no remeshing (blue), remeshing every 0.05 $\tau_0$ (green), and remeshing every 0.01 $\tau_0$ (red). On the right are shown the corresponding shape progressions. In the case of no remeshing, bigger and highly skewed triangles are found in the neck region. These elongated triangles produce an artificial effective increase in the bending stiffness, making neck formation unfavorable. }
  \label{fig:SI-Remesh}
 \end{figure}

{Remeshing is needed to keep the triangles in the mesh as regular as possible while the vesicle grows and deforms. Without remeshing, triangles become skewed and there is an increase in the effective bending stiffness (Fig. \ref{fig:SI-Remesh}). Remeshing is done periodically every $5\cdot 10^{-4}$-$5\cdot 10^{-2} \text{time}/\gamma$,  which we found to be the minimal frequency that ensured the mesh remained regular over the course of the simulation for each set of parameters. Remeshing is done using local mesh operations of edge flips, vertex shifts, edge splits, and edge collapses, as described in \cite{Botsch2010}}

{The reference surface is renormalized to the new number of triangles after remeshing. Each remeshed vertex is projected onto the old mesh to avoid altering the shape of the surface. For convex regions, introducing vertices slightly decreases the volume. The area and volume forces ensure that any changes in surface area and volume introduced by remeshing are rapidly equilibrated.}
 
  \subsection{Parameters}
To estimate the dimensionless parameters relevant for a prebiotic scenario, we use the following values: $\mu=0.8\cdot 10^{-3} \,\text{kg}/\text{m}\cdot\text{s}$ for the water viscosity at room temperature, $B=10 \,k_\text{B}T=4\cdot 10^{-20}\,\text{J}$ \cite{Rawicz2000} for the bending stiffness, $K\Delta P=10^{-7}$--$10^{-5}\,\text{m/s}$ for the membrane permeability, $\gamma=0.5\, \text{s}^{-1}$ for the growth rate and $\abs{c_0} =10^6$--$10^8\,\text{m}^{-1}$ \cite{Kamal2009,Kooijman2005} for the magnitude of spontaneous curvature. The growth rate is taken from a synthetic experimental system in which prebiotic conditions were explored \cite{Chen2004b}. For the permeability we note that different groups have measured membrane permeabilities ranging between $10^{-10}$--$10^{-3}\,\text{m/s}$ \cite{Sacerdote2005,Olbrich2000, Huster1997,Jansen1995}. (We consider only a subset of this wide range on the basis that, even if simple lipids are more permeable, simple cells could probably not have sustained high osmotic pressures. The subset of values we consider corresponds to the regime of membrane-driven growth.) Given these values, we estimate the  relevant parameter ranges in a prebiotic scenario to be $\Pi_1 \sim 0.01$--$1$ and $|\Pi_2| \sim 0.1$--$100$ (Table \ref{tab:dimlessparam3D}).
  \begin{table}[hb!]
 \centering
 \begin{tabular}{|l|c|c|}
 \hline
 &\quad \textbf{Biologically relevant values} \quad&\quad  \textbf{3D Simulation values} \quad\\ \hline
 \quad $\Pi_1$   \quad    & 0.01\text{--}1      & 0.01\text{--}0.5     \\
 \quad $\Pi_2$  \quad     & 0.1\text{--}100       & 0\text{--} 5   \\
 \hline
 \end{tabular}
 \caption{Comparison of the dimensionless parameters relevant in a prebiotic scenario and those used in 3D simulations. $\Pi_2$ is listed in absolute values independent on the curvature direction.}
 
 \label{tab:dimlessparam3D}
 \end{table}

 {We perform simulations over the range $\Pi_1=0.01$--$0.5$ and $\Pi_2=-2.5$--$5$ by varying the growth rate, permeability, bending stiffness, viscosity and spontaneous curvature. Although the exact value of these parameters is not important and the dynamics are controlled solely by the dimensionless parameters, we found that for the simulations to be done in a feasible time, the parameters were better kept in the following limits: $B = [10$ -- $1000] \epsilon_0$, $\gamma = [10^{-5}$ -- $10^{-3}] \tau_0$, $K\Delta P = [10^{-4}$ -- $10^{-2}] \sigma/\tau_0$, $\mu= [1$ -- $1000] m_0/\sigma\tau_0$, $|c_0| = [0.05$ -- $5] \sigma^{-1}$, where $m_0$, $\sigma$,$\tau_0$ and $\epsilon_0$ are the system units for mass, length, time and energy.}
  
 {$R_x$ and $R_i$ were kept constant in each column in the phase diagram of the Main Text (see Table \ref{tab:params} for their values). For each simulation, the vesicle was initialized as a sphere of radius equal to $2R_i$ and is discretized using a mesh consisting of about 7,000 triangles of edge length $l_0=0.5\sigma$. Given that our model cannot currently capture fusion and division, simulations were run up until material from opposite sides of the vesicle membrane came close to touching. The final configurations are therefore interpreted as vesicle shape just prior to fusion or division.}
   \begin{table}[h]
 \centering
 \begin{tabular}{|l|c|c|c|c|}
 \hline
 \quad \textbf{$\Pi_1$} \quad&\quad 0.02  \quad&\quad  0.05  \quad&\quad 0.15  \quad&\quad  0.25\\ \hline
 \quad  $R_i/l_0$   \quad    & \quad 5  \quad&\quad 10    \quad&\quad 8      \quad &\quad 5 \\
 \quad $R_x/l_0$  \quad     & \quad 232   \quad&\quad  200  \quad&\quad 54  \quad&\quad 20 \\
 \hline
 \end{tabular}
 \caption{Values used for $R_i$ and $R_x$ given as a function of the triangle edge $l_0$ for different values of the dimensionless parameter $\Pi_1$.}
  \label{tab:params}
 \end{table}
 
 {We chose $k_a$ in the limit where bending deformations are favored over stretching or compression, i.e.~$k_a l_0^2 \ge B$, which is consistent with the behavior observed for biological membranes. Since $B$ adopts different values in the simulations, $k_a$ is a function of $B$. For all our simulations $k_a l_0^2/B$ was chosen in the range 0.5--10, being $l_0$ the edge length of the mesh triangles. (The lower limit was only used in the case $\pi_1 =0.02$, which was highly constrained since   a large value of $B$ was needed to keep the simulation time short but $k_a$ needed to be sufficiently small to avoid numerical instabilities. In those cases we still confirmed that the membrane did not stretch by over 2\%.)}
 
 {A non-zero temperature was used to check the robustness of our results to noise. However, since mesh elements need to be smaller than the noise wavelength for numerical stability, the temperature was fixed at an artificially low value to allow for larger triangles and faster simulation. (Because thermal fluctuations occur on a much shorter scale than bending deformations, we do not expect this approximation to have a significant effect on the growth dynamics.)}
 
 \subsection{Comparison with local hydrodynamic formulation}
 
\begin{figure}[h!]	
	 \includegraphics[height=3in]{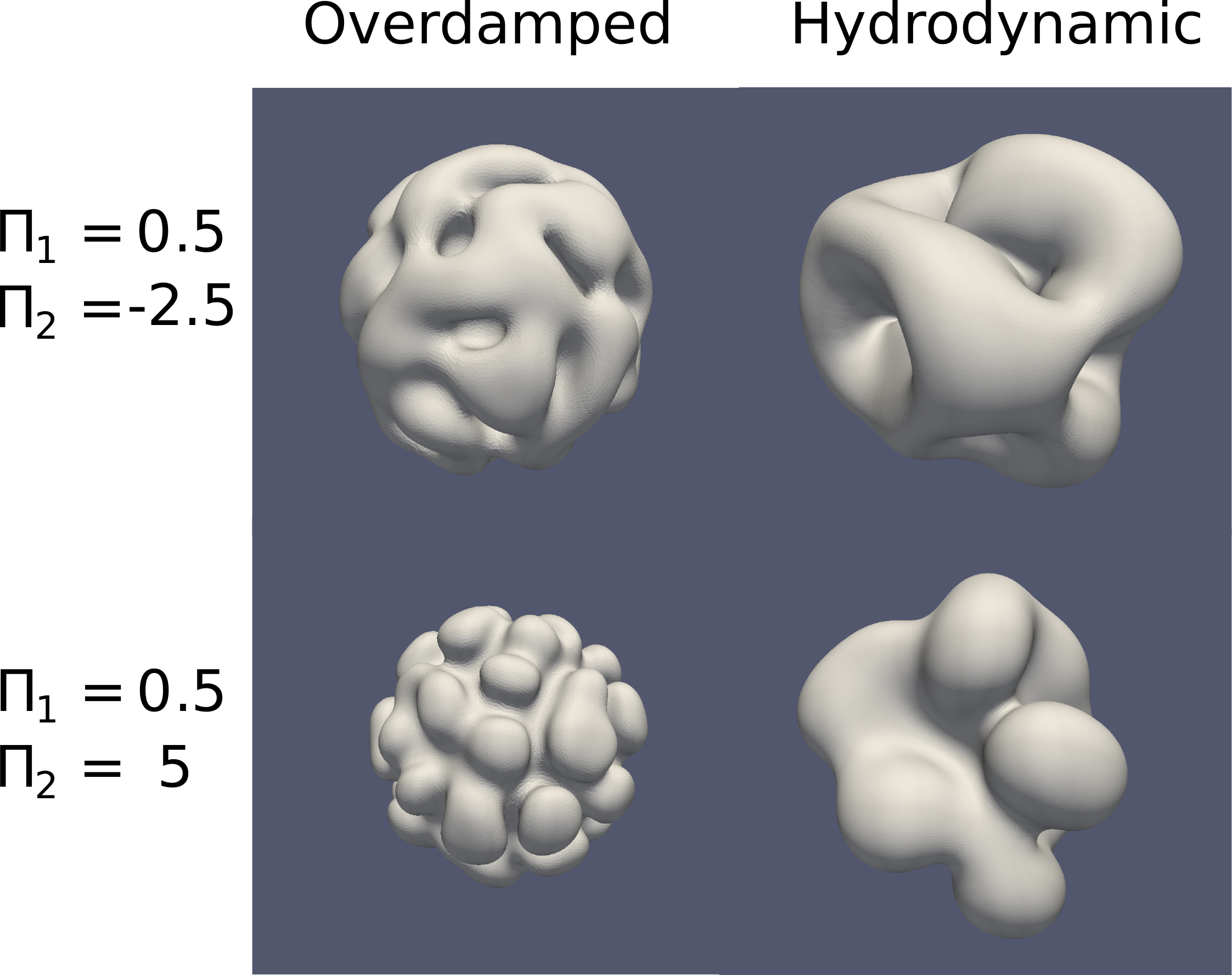}
         \caption{Comparison between immersed boundary and overdamped simulations in two representative cases.}
\label{fig:SI-ibvsoverdamped}
        \end{figure}

 We have compared immersed boundary simulation results to the overdamped simulations of the Main Text (Fig.~\ref{fig:SI-ibvsoverdamped}). We find that the incorporation of fluid dynamics increases the lengthscale of membrane tubules and invaginations and leads to more large-scale crease and folds.
 
Table \ref{tab:table3D} contains descriptions of all parameters along with their numerical values. The growth rate in these simulations ranges from $\gamma=0.6$--$100\, s\textsuperscript{-1}$, whereas the permeability ranges from $K=0$--$0.05\, \text{m/(Pa}\cdot\text{s)}$, with units of flow rate per area times pressure. For the parameters governing membrane elasticity, the bending modulus is chosen to be $B=1 \times 10^{-19}\, \text{J}$ and the stretching coefficient is chosen within the range $k_a=10^{-5}$--$10^{-4} \,\text{J}/\text{m}^2$. Because many biological membranes are to good approximation locally area-preserving, $k_a$ is chosen sufficiently large so that the total vesicle area does not change by more than a few percent during our simulations, but not so large that an impractically small time step is required. We take the fluid density $\rho$ to be $10^3\, \text{kg}/\text{m}^{3}$ and the fluid viscosity $\mu$ to be $10^{-4}\,\text{Pa}\cdot s$. (Using this reduced viscosity compared to water's viscosity of $10^{-3}\,\text{Pa}\cdot s$ allows for faster membrane equilibration and less computation time.) Together with the approximate vesicle radius of $3\, \mu \text{m}$ and velocities on the order of $10^{-4}\,\text{m/s}$ observed during simulations, these fluid parameters result in an approximate Reynolds number of $10^{-2}$. To compute the dimensionless parameters $\Pi_1$ and $\Pi_2$, we make use of the effective membrane thickness $l$ obtained by dimensional analysis of the elastic moduli: for Young's modulus $Y$, $k_a = Y l$ and $B = Y l^3/12$, which leads to $l = \sqrt{12B/k_a}$. 

\begin{table}[h!]
  \centering
  \caption{Caption for the table.}
  \label{tab:table3D}
  \begin{tabular}{cccc}
    \toprule
    \specialrule{1pt}{0pt}{0pt}
    Symbol & Definition & Value & Units\\
    \midrule
    \specialrule{1pt}{0pt}{0pt}
    $\gamma$ & Growth rate & 0.6 - 100 & s$\textsuperscript{-1}$\\
    $K$ & Permeability & 0-0.05 & $\text{m/(Pa}\cdot\text{s)}$\\
    $\mu$ & Fluid viscosity & $10^{-4}$&$\text{Pa}\cdot \text{s}$\\
    $\rho$ & Fluid density & $10^3$ &$\text{kg}/\text{m}^{3}$\\
    $B$ & Bending modulus &$10^{-19}$ & \text{J}\\
    $k_a$ & Bulk modulus & $10^{-5}$--$10^{-4}$ & $\text{J}/\text{m}^2$\\
    $R$ & Vesicle radius &3 & $\mu \text{m}$\\
    \bottomrule
    \specialrule{1pt}{0pt}{0pt}
  \end{tabular}
\end{table}

 \section{2D  Model}
 
  In 2D the spontaneous curvature does not play a critical role and the full space of morphologies can be captured by the single dimensionless parameter $\Pi_1$ (Fig.~\ref{fig:SI-Fig2D}a)
All initial conditions are observed to evolve toward one of the following modes: symmetric division, external budding, or internal vesiculation. Which of these modes is realized depends on the degree of imbalance between surface to volume growth.

Upon observing many replication cycles we conclude that these three modes map onto two essential periodic steady states: vesicles that self-replicate with various degrees of symmetry, giving rise to new generations having very small dispersion in size (Fig.~\ref{fig:SI-Fig2D}b), and vesicles that reproduce via internal budding of smaller vesicles (Fig.~\ref{fig:SI-Fig2D}c).

In either periodic steady-state, homeostatic behavior arises naturally, as the newly generated vesicles always have the same initial size (Fig.~\ref{fig:SI-Fig2D}b,c). In the case of symmetric division, vesicles that grow into cigar shapes display accurate size control when the permeability and growth rate are such that both the perimeter and area double simultaneously (Fig.~\ref{fig:SI-Fig2D}b). This ensures that the two daughter vesicles have the same size as the mother vesicle. Since division occurs when the vesicle doubles its size, the time between divisions is $\tau=\ln{2}/\gamma$.

 \begin{figure}
  \centering
\includegraphics[bb=0 0 2320 880, width=0.8\textwidth]{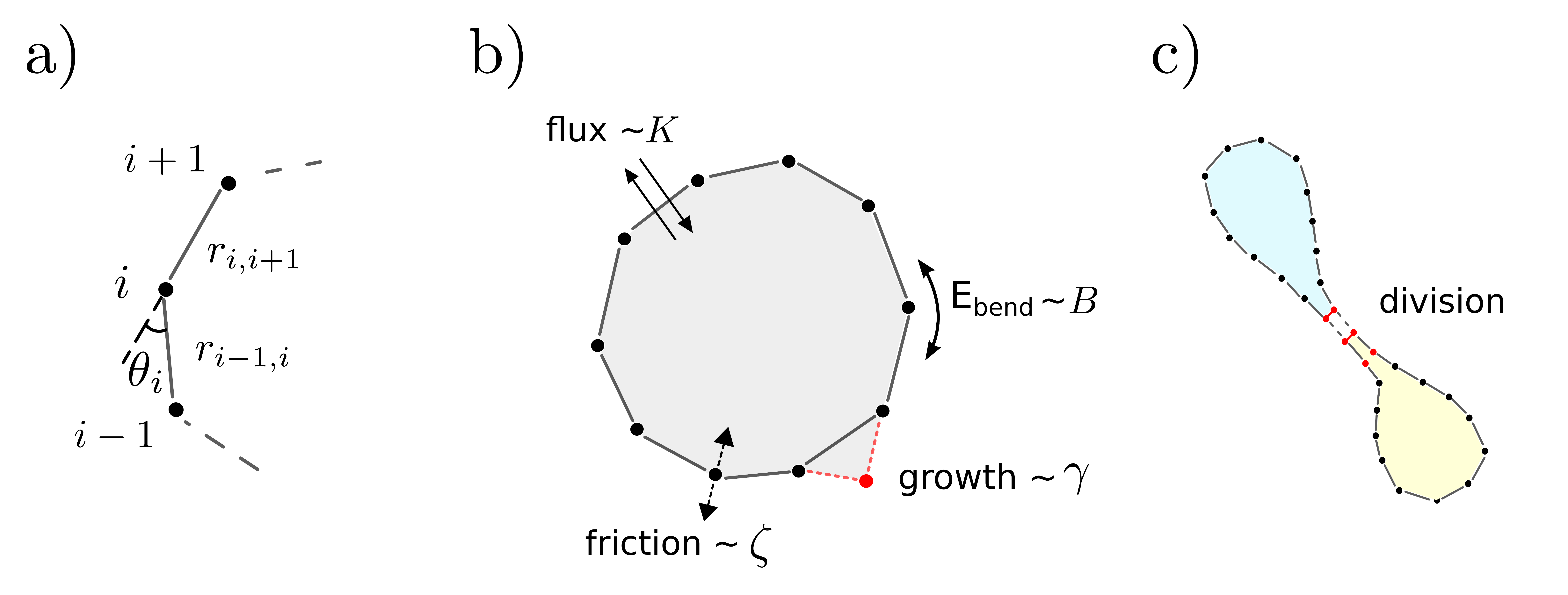}

\caption {Schematic of the system used. a) A section of the spring network. b) The vesicle perimeter grows when a new node is added (which occurs with rate $\gamma$), and the area evolves according to the permeability. c) Division occurs when two non-neighbor sections approach each other more closely than a threshold value.}
\label{fig:SI-Fig2D2}

 \end{figure}

   \begin{figure}
 \centering
  \includegraphics[bb=0 0 2192 1228, width=0.95\textwidth]{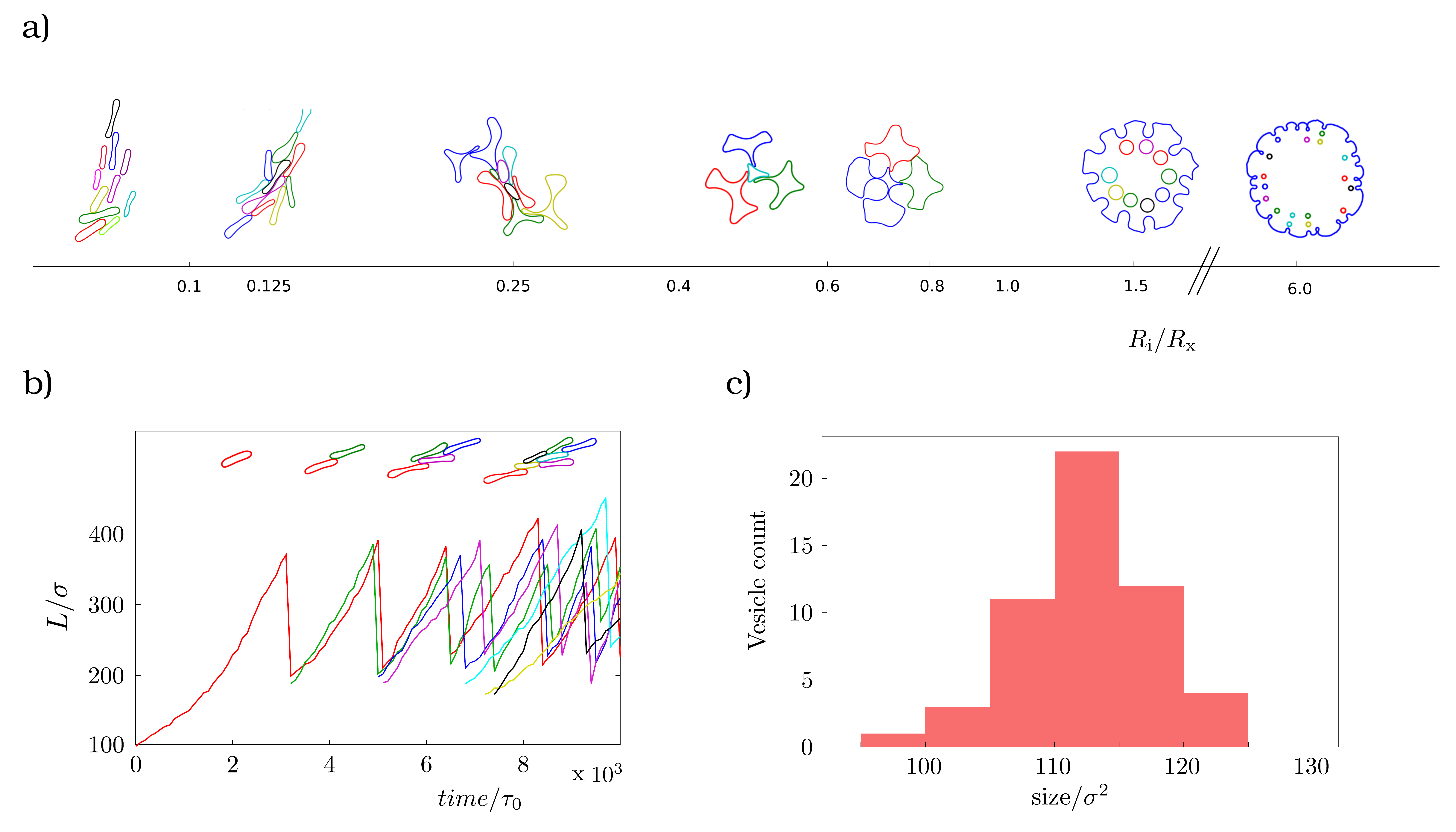}

\caption {a) Morphospace for vesicles in 2D as a function of the dimensionless permeability $\Pi_1 = R_i /R_x $, b) Symmetric division of cigar shape vesicles: perimeter as a function of time and homeostasis during multiple cycles of growth
and division, c) During internal vesiculation, a periodic steady state is reached in which division occur symmetrically with very little size dispersion.}
\label{fig:SI-Fig2D}
 \end{figure}

 The dynamics of the vesicle node positions $\mathbf{r}_i$ are simulated by a hybrid molecular dynamics algorithm in which Monte Carlo moves are introduced to account for growth and division. For the 2D model, we assume local hydrodynamics, so that the node positions follow from the solution of a Langevin equation $m\mathbf{\ddot{r}_i}=\mathbf{f}_i-\zeta\mathbf{\dot{r}}+\sqrt{2\zeta k_\text{B}T}\mathbf{R}(t)$, with the force on each node given by $\mathbf{f}_i=-\partial V^i_{TOT}/\partial\mathbf{r}_i$, where $V^i_{TOT}$ includes the stretching and bending energies of the vesicle and steric energy to avoid the overlapping of nodes. 
 
 \begin{equation}
V^i_{TOT}=V^i_{\text{bond}}+V^i_{\text{bend}}+V^i_{\text{rep}}.
\label{eq:pottot}
\end{equation}
 
 The bond potential on each node has a contribution from the interaction with the two neighboring nodes, so that $V^i_{bond}=V^{i-1,i}_{bond}+V^{i,i+1}_{bond}$, with the spring potential
 
  \begin{equation}
V^{i,j}_{\text{bond}}=\frac{1}{2}\kappa_{\text{bond}}(r_{ij}-b_0)^2,
\label{eq:potbond}
\end{equation}
where $r_{ij}$ is the distance between nodes $i$ and $j$ and $b_0$ is the rest length.
 
The bending energy, in turn, restricts the local curvature $c_i$ on each node according to a harmonic relation with rest curvature $c_0$ (eq.\ref{eq:potbend}). The curvature is defined as the change in angle per unit length between the previous node and the next node, $c_i=\frac{2\theta_i}{r_{i-1,i}+r_{i,i+1}}$ (Fig. \ref{fig:SI-Fig2D2}).

  \begin{equation}
V^i_{\text{bend}}=\frac{1}{2} \kappa_{\text{bend}}(c_i-c_0)^2.
\label{eq:potbend}
\end{equation}

Finally, repulsion between pairs of nodes is modeled via a WCA potential \cite{Weeks1971}:
 \begin{equation}
V^i_{\text{rep}}=\sum_{\substack{j=1{}\\j\ne i}}^n V_{\text{WCA}}(r_{ij}),
\end{equation}

 \begin{equation}
V^i_{\text{WCA}}(r)=\left\{\begin{array}{cr}
4 \epsilon_{0} \left[ \left( \frac{\sigma}{r}\right) ^{12}-\left( \frac{\sigma}{r}\right) ^{6}+\frac{1}{4}\right] &; \quad r\leq r_\text{c} \\
0 &; \quad r>r_\text{c},
\end{array} \right.
\label{eq:wca}
\end{equation}
where $r_\text{c}=2^{1/6}\sigma$, $\sigma=0.8 b_0$ and $j$ runs over all particles for $i\ne j$.

 The overdamped dynamics are formulated in terms of the friction coefficient $\zeta$ and diffusivity $\sqrt{2\zeta k_\text{B}T}$, with $\mathbf{R}(t)$ a delta-correlated stationary Gaussian process having zero mean and thereby satisfying fluctuation-dissipation balance. Fluid incompressibility and permeability imply that the evolution of the vesicle target area follows $\dot{A_\text{T}}= LK \Delta P$, where $\Delta P$ is the osmotic pressure on the nodes, which are given by $\Delta P_{i}= P_\text{osm}$ which is assumed to be constant. The algorithm consists of a two step Verlet scheme in which the node positions are first updated according to the forces and then the vesicle area is updated to equal $A_\text{T}$. Membrane growth is implemented by introducing new nodes with a probability that depends on the current number of nodes, so that growth evolves exponentially with $\dot{n}=\gamma n$. Finally, vesicle division is allowed when two non-consecutive sections of the filament are closer than a threshold value (Fig.~\ref{fig:SI-Fig2D2}). In order to reach the periodic steady-state, we run the simulations until vesicles undergo several cycles of division corresponding to 15--20 new vesicles.

Note that for a two-dimensional vesicle, pressure has units of force/length and bending stiffness has units of $\text{force x length}^2$. Moreover, we refer to $K$ as a permeability although it has units of (time x $\text{length}^2$)/mass in 3D and (time x $\text{length}$)/mass in 2D, while permeability is usually assigned units of length/time or $\text{length}^2$. It should properly be referred to as a fluid resistance or a scaled permeability.

 \begin{table}[]
\centering

\begin{tabular}{|l|c|c|}
\hline
&\quad \textbf{Biologically relevant value} \quad&\quad  \textbf{2D Simulation values} \quad\\ \hline
\quad $\Pi_1= R_i/R_x$   \quad    & $0.01\text{--}1$      & $0.05\text{--}6$     \\ \hline
\end{tabular}
\caption{Comparison of the dimensionless parameters relevant in a prebiotic scenario and the ones used in 2D simulations}
 
\label{tab:dimlessparam}
\end{table}

 {We set the units of energy, length, mass and time in our simulations equal to the characteristic energy, size, mass and diffusion
time for a node: $\epsilon_0$, $l$, $m_0$ and $\tau_0$ respectively, with $l$ also corresponding to the membrane thickness. In the system units, the parameters we used take the following values: $\sigma=0.8 l$, $b_0=l$, $k_a'=5000 \epsilon_0/l^2$, and $B'=50 \epsilon_0 l$  (Table \ref{tab:sysunits}).
To simplify the exploration, we keep the initial vesicle perimeter and the thermal energy constant, with $L_0=n_0 b_0=100 l$ that corresponds to $R\sim 16 l$, and $\text{k}_BT/\epsilon_0=1$, and the spontaneous curvature equal to the curvature of the initial vesicle $c_0=4\pi^2/L_0$. We find outcomes describing experimental results in the range $\pi_1= 0.01 -1$ by varying the growth rate, the permeability, and the viscosity within the following ranges:  $\gamma=[5\cdot10^{-4}-10^{-2}]\, \tau_0^{-1}$, $K'=[10^{-4}-10] \,\tau l/m_0$, and $\mu=[1-500] \,m_0/l\tau_0$.
These parameters can be assigned physical values by setting the system to room temperature T = 300 K and noting that the typical thickness of a lipid bilayer is around 5 nm, the mass of a typical phospholipid is about 660 g/mol, and the phospholipid area density is on the order of $\eta=0.8\, l^{-2}$. The units of our system can then be assigned as follows: $l=5$ nm, $m_0 =4\cdot 10^{-23}$ kg, $\epsilon_0= 4.14\cdot 10^{-21}$ J, and $\tau_0 = l \sqrt{m_0 /\epsilon_0} = 0.49$ ns (Table \ref{tab:sysunits}).}

 \begin{table}[]
\centering

\begin{tabular}{|lll|ll|ll|}
\hline
&\textbf{Systems units} &&  \textbf{Coefficients} && \textbf{Parameters} &\\ \hline
& $l=b_0$       & 5nm                 &  $\sigma$  & 0.8$l$ & $\gamma$ & $5\cdot10^{-4}-10^{-2} \tau_0^{-1}$\\
&  $m_0$         & $4\cdot 10^{-23}$ kg  & $L_0$ & 100 $l$ &  K' & $10^{-4}-10 \tau l/m_0$ \\
&  $\epsilon_0$   & $4.14\cdot 10^{-21}$ J &$K_a'$ & 5000 $\epsilon_0/l^2$ & $\zeta$ & 1-500 $m_0/\tau_0$ \\
&  $\tau_0$       & 0.49 ns               & $B'$  &  50$\epsilon_0 l$ &&\\ 
&  &  & $R$ & $16l$&& \\ \hline
\end{tabular}
\caption{System Units, coefficients and parameters used in 2D simulations}

\label{tab:sysunits}
\end{table}

\FloatBarrier
 
\bibliographystyle{ieeetr}
\bibliography{biblio}